\documentclass{acm_proc_article-sp}

\usepackage[hyphens]{url}
\usepackage{booktabs}
\usepackage{subfigure}
\usepackage{amsmath}
\usepackage{algorithm}
\usepackage[noend]{algpseudocode}
\usepackage{graphicx}
\usepackage{amsfonts}
\usepackage{amsmath}
\usepackage{multirow}
\usepackage{balance}

\begin{document}

\title{Followers or Phantoms?\\An Anatomy of Purchased Twitter Followers}
\author{Anupama Aggarwal, Ponnurangam Kumaraguru \\
CyberSecurity Education and Research Center (CERC) \\
Indraprastha Institute of Information Technology, Delhi\\
\{anupamaa, pk\}@iiitd.ac.in }

\maketitle

\begin{abstract}
Online Social Media (OSM) is extensively used by  contemporary Internet users to communicate, socialize and disseminate information. This has led to the creation of a distinct online social identity which in turn has created the need of online social reputation management techniques. A significant percentage of OSM users utilize various methods to drive and manage their reputation on OSM. This has given rise to underground markets which buy/sell fraudulent accounts, `likes', `comments' (Facebook, Instagram) and `followers' (Twitter) to artificially boost their social reputation. In this study, we present an anatomy of purchased followers on Twitter and their behaviour. We illustrate in detail the profile characteristics, content sharing and behavioural patterns of purchased follower accounts. 

Previous studies have analyzed the purchased follower markets and customers. Ours is the first study which analyzes the anatomy of purchased followers accounts. Some of the key insights of our study show that purchased followers have a very high unfollow entropy rate and low social engagement with their friends. In addition, we noticed that purchased follower accounts have significant difference in their interaction and content sharing patterns in comparison to random Twitter users. We also found that underground markets do not follow their service policies and guarantees they provide to customer. Our study highlights the key identifiers for suspicious follow behaviour. We then built a supervised learning mechanism to predict suspicious follower behaviour with 88.2\% accuracy. We believe that understanding the anatomy and characteristics of purchased followers can help detect suspicious follower behaviour and fraudulent accounts to a larger extent. 
\end{abstract}


\section{Introduction}
\label{sec:introduction}
\subsection{Research Aim and Motivation}
\label{subsec:aim}
Online Social Media(OSM) like Twitter, Facebook, YouTube and Instagram are being used by Internet users to interact and spread information by enabling them to maintain their online identity. This online identity is based on content sharing and interaction patterns. To boost the reputation and popularity of their online social profiles, users utilize various methods like sharing interesting content, attracting more `likes' and `followers'. This has led to the creation of an underground fraudulent market which promises to boost the reputation of online social profiles by selling `likes', `comments' and `followers'. Recent studies indicate that, selling fake Twitter followers now generates a revenue around \$360 million per year.\footnote{\url{http://cir.ca/news/fake-twitter-followers}} The cost of buying 1,000 Twitter followers can be as low as \$2 which has further enabled these underground markets to perpetrate spam on Twitter. Most of the underground markets claim to provide high quality and genuine `comments', `followers' and `likes'. One of the popular services which sells Twitter followers claims - \emph{``you will get followers which are of high quality. Means, you won't see any profiles with egg images or a profile with no tweet at all"}.\footnote{\url{http://www.buyfollowers.co/twitter.html}} This drives a large number of users to use such services who want to increase their social media popularity. Recent articles reveal that even popular and celebrity users on Twitter have admitted to buy followers to look more popular.\footnote{\url{http://www.nytimes.com/2012/08/23/fashion/twitter-followers-for-sale.html}} 

\begin{figure}[h]
  \centering
    \includegraphics[width=0.45\textwidth]{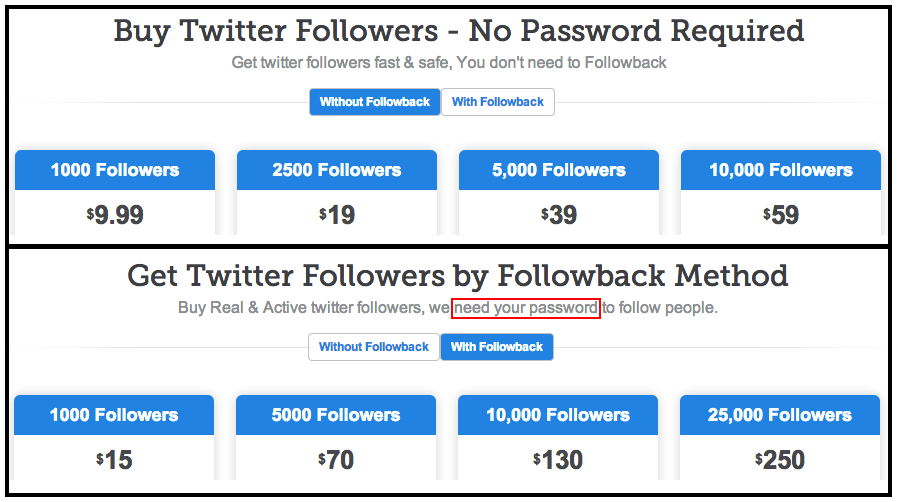}
    \caption{Different purchasing schemes provided by one of the Twitter Follower markets: buyfollowers.co}\label{fig:market}
\end{figure}

Overall, Twitter follower markets provide two popular purchasing schemes -- (i) Without Followback and (ii) With Followback. The primary difference between the two schemes is the requirement of Twitter password by the merchant (i.e., the follower selling websites). In the first scheme, the customer only has to pay for the desired number of followers. When users opt for the second scheme, the merchant asks for the Twitter credentials (password) of the customer users. This enables the merchants to compromise and make the customer user part of fraudulent follower network~\cite{stringhini2013follow}. These compromised accounts can be then used to spread spam and other malicious content. Previous studies reveal that underground follower market contributes to about 10-20\% spam on Twitter~\cite{thomas2013trafficking}. Figure~\ref{fig:market} shows an illustration of the two purchasing schemes discussed above on one of the popular Twitter follower markets (\url{buyfollowers.co}). One of the reasons why these markets have gathered a large customer base is because they claim to provide quality guarantees. Many markets guarantee active and genuine Twitter users as followers. Some markets also guarantee retention of the purchased followers for at least a year. Since the Twitter follower markets generate high revenue by fraudulent methods and the followers provided by them successfully circumvent spam filters; it is important to study the anatomy of these Twitter profiles and find key identifiers of suspicious following behaviour. 

\subsection{Research Contribution}
\label{subsec:contribution}
This study has the following research contributions: 
\begin{itemize}
\item We present an anatomy of the purchased Twitter followers. We characterize the profile attributes and the behavioural features of the purchased followers. We also compare their charecteristics with legitimate users.

\item We identify key indicators to distinguish between suspicious following behaviour from that of genuine Twitter users. We use these identifiers and built a supervised learning mechanism which identifies suspicious following behaviour with an accuracy of 85\%. 
\end{itemize}

Previous studies have detected and analyzed merchants and customers of the underground Twitter follower market~\cite{stringhini2013follow}. Stringhini et al. also detected market victims, i.e., the compromised users that become part of the follow network. Our study however, focusses on the analysis of purchased followers. These followers accounts may not have been compromised by the merchants. Researchers have also explored the at-registration patterns of the purchased followers~\cite{thomas2013trafficking}, that is, the properties of the registered account like email, IP address, number of attempts at CAPTCHA solving etc while registering for the Twitter account with suspicious following behaviour. They studied properties of the accounts put for sale at the time of Twitter account registration and use these features for early detection of fraudulent account registrations. Despite such available techniques, there still exists an underground follower market. In this study, we aim to understand the dynamics of existing Twitter accounts which are sold as `followers' by the underground market. Our study concentrates on understanding the profile characteristics, behavioural and content sharing patterns of purchased Twitter followers. We further use these patterns to automatically detect suspicious following behaviour and present the most discriminative features to do the same.

The rest of the paper is organized as follows -- Section~\ref{sec:related} explains the related work followed by a short introduction to Twitter follower markets in Section~\ref{sec:background}. We explore the anatomy of purchased Twitter followers in Section~\ref{sec:anatomy} and use the discriminative features to detect suspicious follow behaviour in Section~\ref{sec:prediction}. We finally conclude and illustrate some of the possible future work in Section~\ref{sec:future}.

\section{Related Work}
\label{sec:related}
In this section, we present some closely related work focussed on detection of spam and malicious behaviour on Twitter. We also summarize the previous literature on the underground market which enables spam monetization on online social media.

Over last few years, researchers have conducted several studies on online social media and specifically, Twitter. With an increase in use of social media, miscreants have started to spread spam and malicious content on social media~\cite{oscar2005leveraging}. Researchers have proposed several techniques to detect spam and malicious content on Twitter. Some of these strategies involve analysis and detection of sybil nodes~\cite{danezis2009sybilinfer, yu2006sybilguard}. Previous studies have also used Twitter based charecteristics to identify features which can be helpful to detect spam users~\cite{benevenuto2010detecting, lee2010uncovering, stringhini2010detecting}. URL based methods have used blacklist lookup and URL redirection chains to detect spread of malicious content~\cite{lee2012warningbird, thomas2011design}. Researchers have also shown that spammers use compromised accounts to spread spam on Twitter~\cite{gao2010detecting}. Various methods like finding common content sharing patterns, modelling user behavior and detecting bot-like tweeting activity have been used to detect compromised accounts~\cite{egele2013compa, ghosh2012understanding, thomas2011suspended}.

Recent studies have shown that miscreants use several strategies to monetize spam and other malicious activities~\cite{levchenko2011click}. There exists a huge underground market which sells specialized services and products like fraudulent accounts~\cite{thomas2011suspended, thomas2013trafficking}, CAPTCHA-solving~\cite{motoyama2010re}, pay-per-install~\cite{caballero2011measuring} and writing fake reviews or website content~\cite{motoyama2011dirty, wang2012serf}. Such underground markets are a threat to quality of service and is generating a revenue of about \$360 million  per year from sale of fake Twitter followers.\footnote{\url{http://cir.ca/news/fake-twitter-followers}} 

Twitter follower market is one of the most popular underground markets. Users attempt to gain followers in order to boost their popularity~\cite{cha2010measuring}. Researchers have modelled suspicious following behaviour by identifying difference in follow pattern from the majority~\cite{jiang2014detecting}. Researchers have previously studied how underground markets operate and understand the dynamics of merchants and customers. They studied the unfollow dynamics of the victim accounts whose credentials are compromised by the merchants~\cite{stringhini2013follow}. However, our study focusses on the analysis and characterization of purchased Twitter followers. Previous studies have characterized the registration time properties of purchased accounts like the email address used, originating IP address and time taken for creation of accounts. Researchers investigated the economy of follower markets and estimated the revenue they generate by selling fraudulent accounts and services. Researchers used these at-registration-time properties of the accounts sold by the merchants to identify features which can be used to detect fraudulent accounts as and when they are created by the merchant~\cite{thomas2013trafficking}. Despite such techniques, there exists a large underground market which promotes the sale of Twitter followers. In order to better understand the dynamics of purchased followers and deter such practices, we present an anatomy of purchased follower accounts. Our study explores the profile characteristics, content sharing patterns and behavioural features of purchased follower accounts. 

\section{Twitter Follower Market}  
\label{sec:background}
\subsection{Purchase Schemes}
\label{subsec:scheme}
In this section we briefly explain how Twitter follower market operates. The underground market of Twitter followers constitutes more than two-dozen services which generate an annual revenue of \$360 million by the sale of followers. These markets sell followers in bulk and have various purchase schemes. Figure~\ref{fig:market} illustrates two popular purchasing schemes of the follower markets. The cost of bulk followers may differ from one market to another, however, most markets offer the following two purchasing schemes -

\paragraph{Without Followback}
In this scheme, the customer has to provide the Twitter handle (@username) for which he wants to purchase followers and select the number of bulk accounts. The user himself does not need to follow back other users to gain followers. This purchasing scheme is convenient to use, since the user does not need to provide his Twitter credentials to the merchant. This enables the customer to purchase bulk followers for any Twitter handle. This has been exploited by hoaxers in past where they spammed a popular news website's Twitter account with 75,000 fake followers.~\footnote{\url{http://www.dailydot.com/technology/socialvevo-swenzy-fake-twitter-followers-spam-attack/}} In this study we purchase Twitter followers via this scheme to create our ground truth dataset. We use multiple Twitter accounts to gain followers from different services. We describe our dataset in more detail in Section~\ref{subsec:dataset}.
\paragraph{With Followback}
In this scheme, the customer has to provide the Twitter credentials (password) of the account for which he wants to gain bulk followers. This allows the merchant to include customer in the fake follower network by making his account follow other accounts and customers. In this scheme, the customer's account is at the risk of being compromised since merchant gets the Twitter account password. Previous studies have shown that such compromised accounts are used to spread malicious URLs and tweets promoting the merchants~\cite{stringhini2013follow}. To purchase bulk followers in both the schemes, customers have the option to pay via PayPal, WebMoney or credit cards. After making the purchase, followers are provided to the desired Twitter handle within a couple of hours to few days depending on the choice of purchase scheme and the amount of followers.

\subsection{Service Policies and Guarantees}  
\label{subsec:policy}
The merchants of underground Twitter follower market provide various guarantees to customer at the time of purchase. Many merchants claim to provide authentic followers -- \emph{``Customers who purchase Twitter followers with us are assured to get real followers on time"}.~\footnote{\url{http://buytwitterfollower.org/authentic-services/}} Some merchants also provide retention guarantee where they claim that there will not be any drop in the purchased number of followers -- \emph{``...if you loose any number of followers within a period of 1 year from the date of purchase, we'll refill the page with the lagging followers, at absolutely free of cost"}.~\footnote{\url{http://www.buyfollowers.co/twitter.html}} Many merchants also provide money-back guarantee in case of partial fulfilment of services -- \emph{``If you receive 1 follower less than you ordered we will issue a full refund."}.~\footnote{\url{http://www.followersfortwitter.com/}} Such promising guarantees encourage customers to place bulk orders in order to boost their social reputation. In Section~\ref{subsec:policyanlysis} we illustrate that merchants do not necessarily stick to the guarantees they provide to customers and the purchased followers may not be of high quality or real.

\subsection{Freemium Market}
Apart from the above two purchase policies to gain followers, there also exists a freemium model in Twitter follower underground market. In this model, users have to authorise a third-party Twitter application and in return gets about 60-100 followers. The permissions asked by the application include - \emph{See who you follow, and follow new people}, \emph{Update your profile} and \emph{Post Tweets for you}. Thus, once authorised, the application is able to post promotional tweets about the merchant using the customer user's profile and the user becomes part of the follow-network. This study however focusses only on the premium model where users have to pay to gain followers.

\section{Anatomy of Purchased Twitter Followers}
\label{sec:anatomy}
In this section, we provide an analysis of the characteristics of purchased follower accounts. We describe in detail the profile properties, content sharing and behavioural patterns of purchased followers and highlight their suspicious behaviour.

\subsection{Dataset Description}
\label{subsec:dataset}
For our analysis, we purchased Twitter followers from two merchants. We created two dummy accounts to make purchases from these merchants. To ensure that regular Twitter users do not start following us, we (i) maintained a minimal Twitter profile without a `profile image' or a `bio'; (ii) made the purchase within few minutes after creation of the account. Therefore, we safely assume that all the followers which we gained were from merchants with whom we placed the purchase order. Table~\ref{table:dataset} describes the cost and number of followers we purchased from each merchant.

\begin{table}[h]
\small
\caption{Dataset description of purchased followers from underground market}\label{table:dataset}
\begin{tabular}{@{}lllll@{}}
\toprule
Merchant            & \begin{tabular}[c]{@{}l@{}}Users\\ Purchased\end{tabular} & \begin{tabular}[c]{@{}l@{}}Users\\ Obtained\end{tabular} & \begin{tabular}[c]{@{}l@{}}Date of\\ Purchase\end{tabular} & \begin{tabular}[c]{@{}l@{}}Cost/1000\\ followers\end{tabular} \\ \midrule
buyfollowers     & 1,000                                                       & 1,090                                                      & 10-02-13                                                      & \$9.99 \\
buy1000followers & 10,000                                                      & 11,346                                                     & 04-18-14                                                  & \$1   \\ \bottomrule
\end{tabular}
\end{table}

There exist several Twitter follower markets, however we chose the two mentioned in Table~\ref{table:dataset} because they provided followers at a very cheap rate and offered fast delivery. Notice that the cost of buying followers from buy1000followers.co was one-tenth the price of the first one. We opted for ``without-followback" purchase scheme on both the merchant sites and obtained 12,436 unique followers accounts. Out of these users, 11,760 users have public profiles on Twitter. Some of our analysis in latter sections based on content sharing patterns focusses only on these 11,760 unique users. For all the 12,436 users, we took hourly snapshots of their profile based information which we use in our analysis. Table~\ref{table:description} gives the description of hourly snapshots and number of public profiles from each merchant.

\begin{table}[h]
\centering
\small
\caption{Hourly snapshots and public users}\label{table:description}
\begin{tabular}{@{}lllll@{}}
\toprule
Merchant         & \begin{tabular}[c]{@{}l@{}}\#Hourly\\ Snapshots\end{tabular} & \begin{tabular}[c]{@{}l@{}}Unique\\ Users\end{tabular} & \begin{tabular}[c]{@{}l@{}}Public\\ Users\end{tabular} & Tweets \\ \midrule
buyfollowers     & 1,400                                                     & 1,090                                                  & 902                                                    & 83,936 \\
buy1000followers & 600                                                       & 11,346                                                 & 10,768                                                 & 339,432        \\ \bottomrule
\end{tabular}
\end{table}

For each follower captured in our hourly snapshot, we extract past 200 tweets by that user. Table~\ref{table:description} shows the number of tweets we collected for users from each market. We collect 350,778 tweets to analyse the content sharing pattern by purchased followers.

\subsection{Analysis of Market Service Policy}
\label{subsec:policyanlysis}
Follower markets provide guarantees and service policies to the customers as described in Section~\ref{sec:background}. In this section we describe how markets do not stick to the guarantees they claim.

\subsubsection{Fluctuations and Drop in Followers}
The markets we used to purchase Twitter followers provide `retention guarantee' stating - \emph{We have 1 Year retention guarantee policy which means if you loose any number of followers...we'll refill the page with the lagging followers, at absolutely free of cost}

\begin{figure}[ht]
\centering
\subfigure[Market1 - buyfollowers]{
\includegraphics[scale=.25]{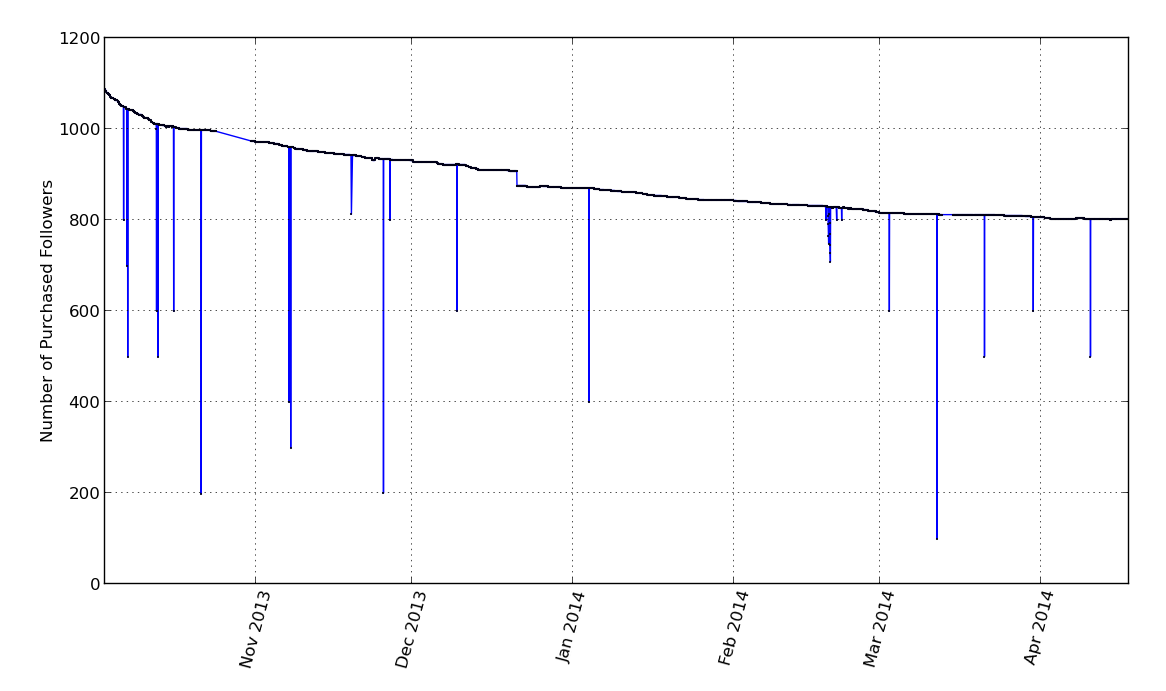}
 \label{subfig:numFollowersExp1}
}
\subfigure[Market2 - buy1000followers]{
\includegraphics[scale=.25]{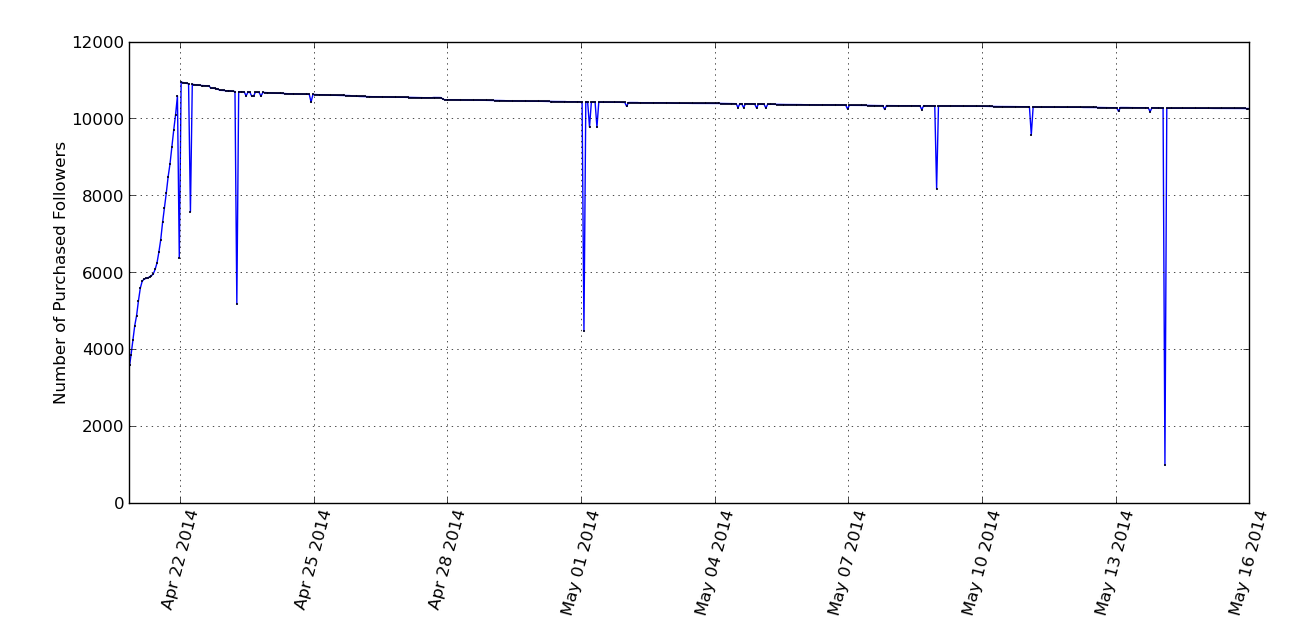}
 \label{subfig:numFollowersExp2}
\label{fig:numFollowers}
}

\caption{Purchased follower fluctuations in different Twitter follower underground markets}\label{fig:numFollowersExp}
\end{figure}

Figure~\ref{fig:numFollowers} shows hourly fluctuations in number of followers since the date of purchase. The figure has various dips where the number of followers reduced drastically. We purchased 1,000 followers from the Market1 (buyfollowers.co), however, after few days the total number of followers were reduced to less than 800 as shown in Figure~\ref{subfig:numFollowersExp1}. We contacted the merchant, but did not get a response. Even after the drop in followers to 800, there were frequent dips with constantly decreasing count of the number of followers. The second market (buy1000followers) exhibits the same pattern in Figure~\ref{subfig:numFollowersExp2}. Note that there were several dips in the number of followers in both markets, however, we did not gain new users as followers. Users from the same set of initially obtained users kept unfollowing and following us back. This also highlights the suspicious behaviour of the purchased followers. 

%
%

To analyse whether the dips in follower count are at a specific time, we measured correlation of follower count with hour of the day. We calculated the \emph{Pearson Correlation Coefficient} (PCC) between the follower count distribution and the corresponding hour of the day of snapshot time. We found that there does not exist any correlation between the follower count and time for both the merchants (PCC = 0.01, 0.009, indicating negligible correlation). 

\begin{figure}[h]
  \centering
    \includegraphics[width=0.46\textwidth]{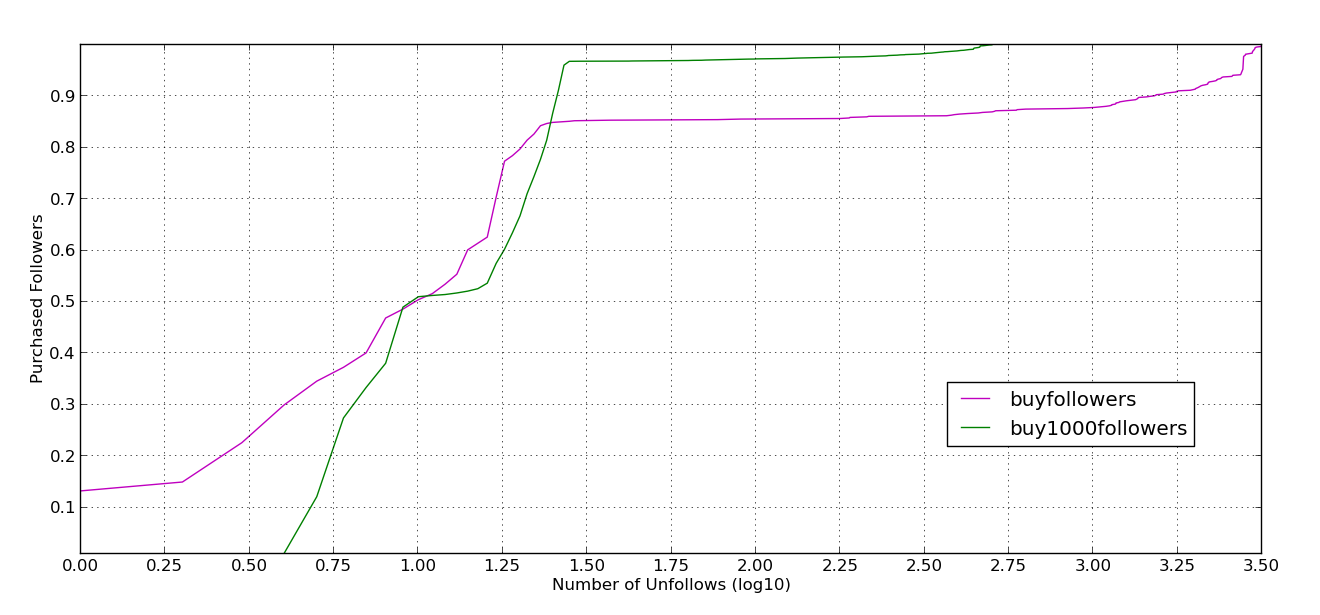}
    \caption{Unfollow Rate of Purchased Followers}\label{fig:UnfollowCount}
\end{figure}

We further analysed the dips in purchased follower count. We noticed that several followers keep unfollowing and following us back. We took hourly snapshots of the purchased followers and found that in the first market, 928 out of 1,090 users unfollowed us one or more times. In the second market, 10,595 out of 11,346 users unfollowed and followed us back.  Figure~\ref{fig:UnfollowCount} shows the unfollow frequency by purchased followers from each market. In the first market, about 85\% users unfollowed us at least 24 times and about 1\% users unfollowed us more than 1,300 times within a span of 7 months consisting of 1,400 hourly snapshots. In the second market, during a span of 400 hourly snapshots, the maximum unfollow rate for a single user was 388. We believe that high unfollow entropy can be useful to detect suspicious following behaviour; we explore this phenomena more in later sections. Also, a large number of dips and spikes in the follower count of a user can put the user under suspicion of having fake followers.  

\subsubsection{Inactive and Suspended Accounts}
\label{subsub:inactive}
The follower markets claim that the purchased followers will be of high quality and will have active users. Stating the service policy of the market from where we purchased followers -- \emph{``you won't see any profiles with egg images or a profile with no tweet at all. They all have complete bio and recent tweets on their timeline."}

\begin{figure}[h]
  \centering
\subfigure[CDF of time (in days) of the last tweet posted by purchased follower accounts]{
        \label{subfig:lasttweet}
        \includegraphics[width=7.2cm]{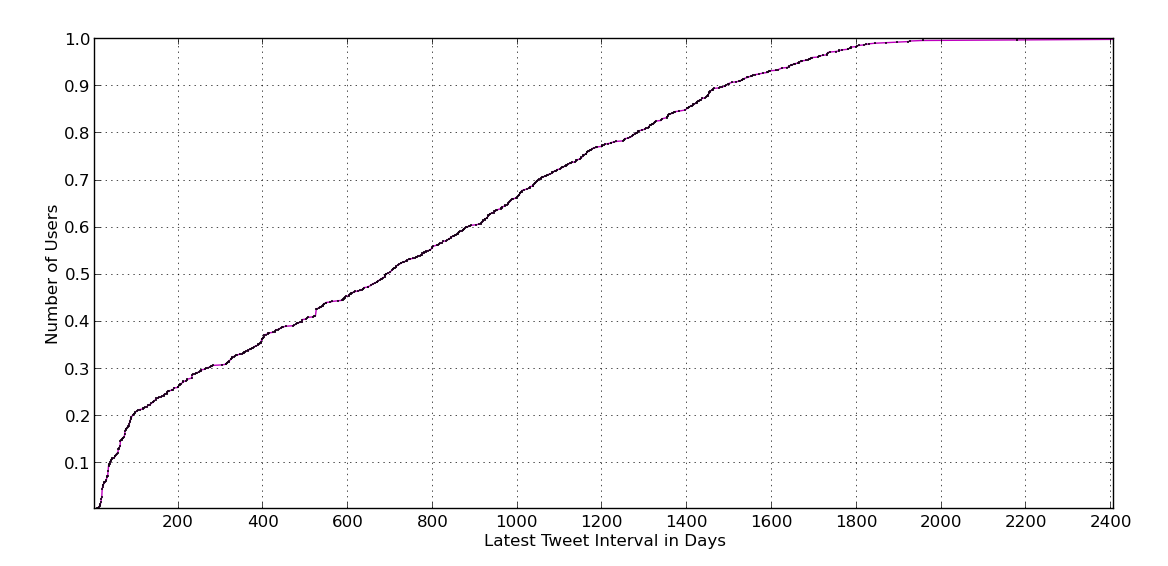} }         
\subfigure[CDF of fraction of Retweets over Tweets posted by the purchased follower accounts]{
        \label{subfig:fraction_RTcombo}
        \includegraphics[width=7.2cm]{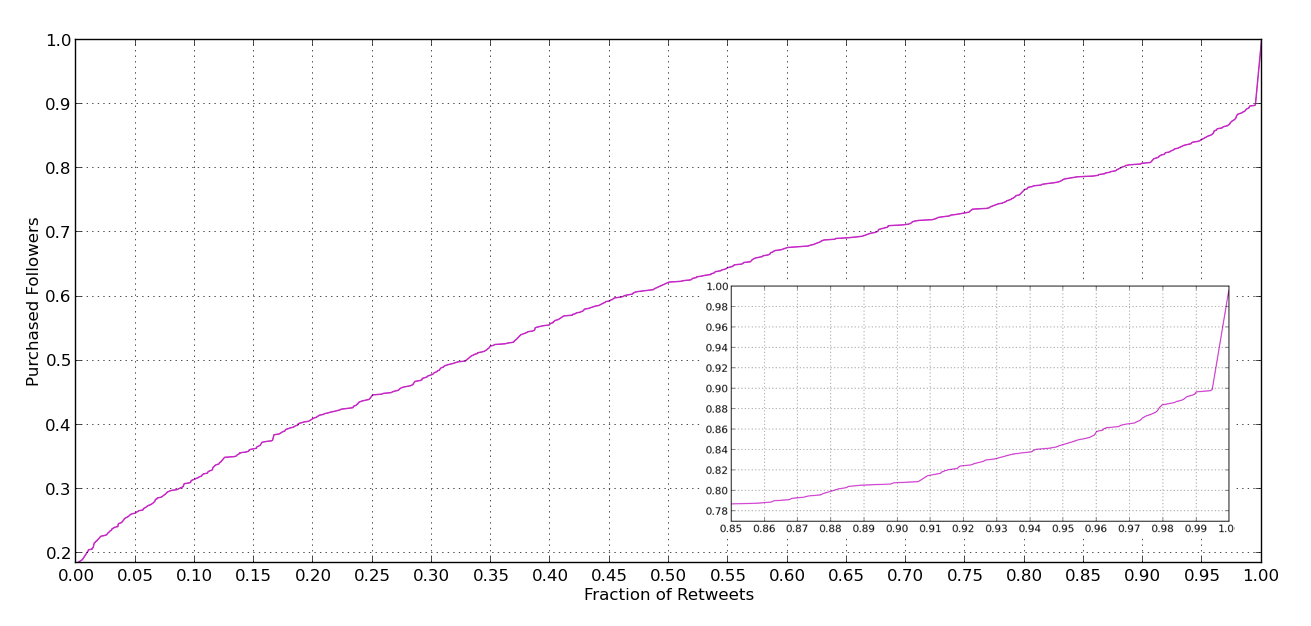}} 

    \caption{Tweet Inactivity of Purchased Followers}\label{fig:tweet_pattern}
\end{figure}

We investigate the quality of purchased followers to find out whether the accounts were active or not. Figure~\ref{subfig:lasttweet} indicates that only 26\% purchased followers had a tweet within past 200 days at the time of our analysis during April 2014. About 45\% users had not tweeted even once in past 2 years. This clearly indicates violation of service policy by the markets and also shows that the purchased followers are of low quality. 

We further analyze the past 200 tweets of all the purchased follower accounts. We found that a large fraction of  users post less original content and only retweet. Figure~\ref{subfig:fraction_RTcombo} shows that 38\% users had less than 50\% original tweets posted on their timeline. Also, 10\% purchased users had more than 99\% of their tweets as only retweets. This shows that the purchased followers do not actively post content themselves but rely on retweeting activity to increase their status count.

Out of the 1,090 followers we obtained from the first market, 55 user profiles were suspended. According to `Twitter rules', accounts are suspended in case of violation of rules.~\footnote{\url{https://support.twitter.com/articles/15790-my-account-is-suspended}} This shows that not all the followers delivered by the merchant were quality profiles. 

We further investigated the profile properties of purchased followers according to the service policy provided by the merchant. We found only 4 users which had the default `egg' profile picture. However, out of the 1,090 purchased followers, 700 profiles did not have a profile description.

\subsubsection{Social Reputation}
\begin{figure}[h!]
  \centering
    \includegraphics[width=0.45\textwidth]{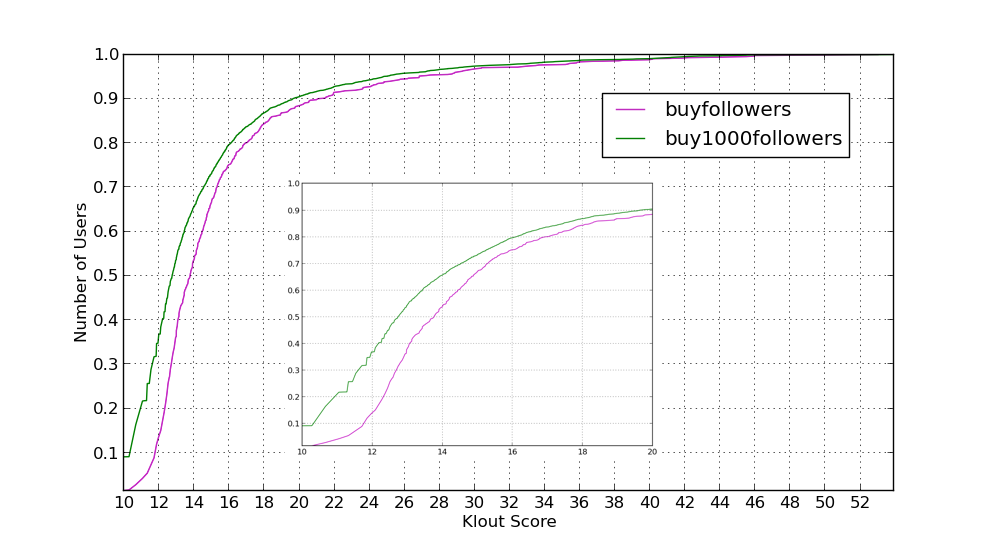}
    \caption{CDF of Klout Score of Purchased Followers}\label{fig:kloutScoreCombo}
\end{figure}

The follower markets claim that the purchased accounts will be real and of high quality -- \emph{``Authenticity of followers is guaranteed"}. To validate the quality of purchased followers we used `Klout'~\footnote{\url{http://www.klout.com}} to determine the social influence. `Klout' is a popular tool to measure influence based on various factors like followers, freinds, retweets and favourites. The average Klout score for the social media users is 40.~\footnote{\url{http://support.klout.com/customer/portal/articles/679109-what-is-the-average-klout-score}} However, as shown in Figure~\ref{fig:kloutScoreCombo}, we found that 90\% of the purchased followers had a Klout score of less than 20. This shows that these accounts do not involve in discussions with other users and have a low influence score.

The above characterization shows that the merchants do not stick to their service policies and guarantees. We also find that the purchased followers are low quality and have a suspicious following behaviour. We use these indicators to build a system which can detect suspicious following behaviour.

\subsection{Network Characteristics of Purchased Followers}
In this section we analyze the network properties of the purchased followers. We look at various factors like the follower and friends count, unfollow entropy, and follower/friends ratio.
\label{subsec:network}
\subsubsection{Follower / Friends Ratio}
We look at the relationship between amount of followers and friends for purchased follower accounts. On Twitter, `followers' of a person are the users which subscribe to the posts of that person, i.e., who `follow' him. The `friends' of a person are the users whom he subscribes to. The average number of followers per existing account is 68 and the average number of friends is 60 on Twitter.

\begin{figure}[h!]
  \centering
    \includegraphics[width=0.55\textwidth]{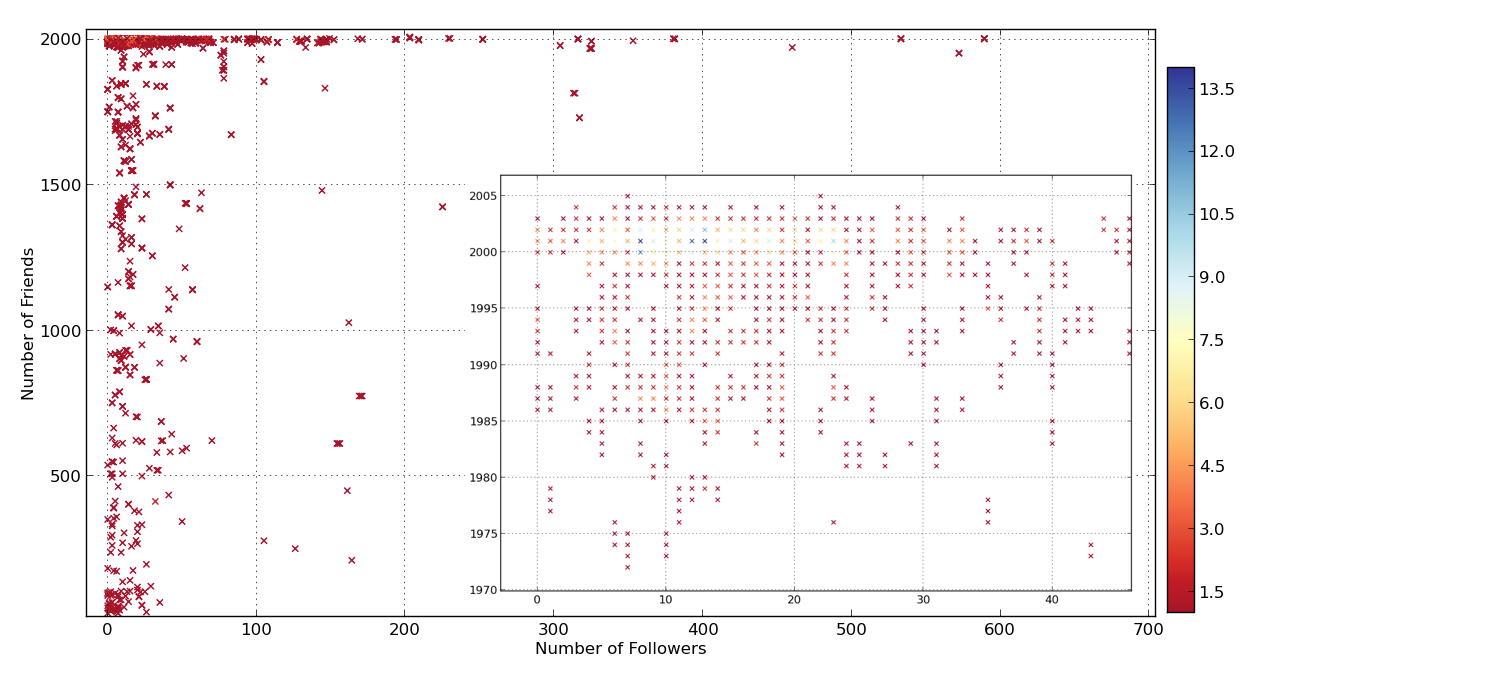}
    \caption{Number of Followers vs Friends of purchased follower accounts}\label{fig:follower-followeeCombo}
\end{figure}

Figure~\ref{fig:follower-followeeCombo} shows a large fraction of purchased follower accounts have low follower count but a very high friends count. It also shows that the difference in the number of followers and friends is large and the purchased follower accounts do not gain a lot of followers themselves. To investigate this further, we look at the Follower-Friends ratio of these accounts.

\begin{figure}[h!]
  \centering
    \includegraphics[width=0.5\textwidth]{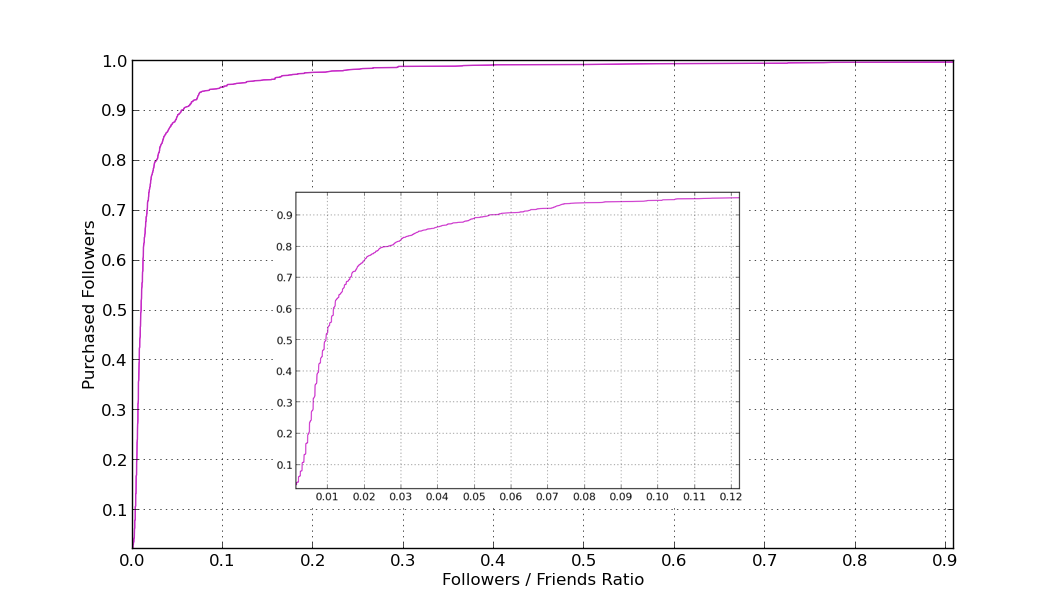}
    \caption{Follower-Followee ratio of purchased follower accounts}\label{fig:follower-followee-ratioCombo}
\end{figure}

We observe in Figure~\ref{fig:follower-followee-ratioCombo} that the follower/friends ratio fits the power law ($\alpha$ = 1.8209, \emph{error} $\sigma$ = 0.029). We observe that 94\% purchased followers have the follower/friends ratio as only 0.1 and none of the purchased followers had more followers than friends. This further strengthens our observation that purchased accounts have a low follower count.

\subsubsection{Unfollow Entropy}

We found that the purchased follower unfollowed a large number of users regularly. To quantify this behaviour, we calculated the \emph{unfollow entropy} of all the purchased followers. We observed each purchased follower over a span of 30 days and collected his hourly followers. We define normalized \emph{unfollower entropy} $H$ for a user $u_n$ as the following 
$$H_{u_n} = - \frac{\Sigma_{i=1}^T p_n(f_i) log(p_n(f_i))}{N}$$
where, $p_n(f_i)$ is the probability that the user $u_n$ will unfollow at time $t_i$. The probability function is defined as
$$p_n(f_i) = \frac{ucount_i}{\Sigma_{i=1}^T ucount_i}$$ where $T$ is the number of days for which we monitor the purchased follower and $ucount_i$ is the number of users he unfollowed on $i^{th}$ day. A higher value of unfollow entropy signifies that the user exhibits a suspicious unfollow pattern.

\begin{figure}[h!]
  \centering
    \includegraphics[width=0.34\textwidth]{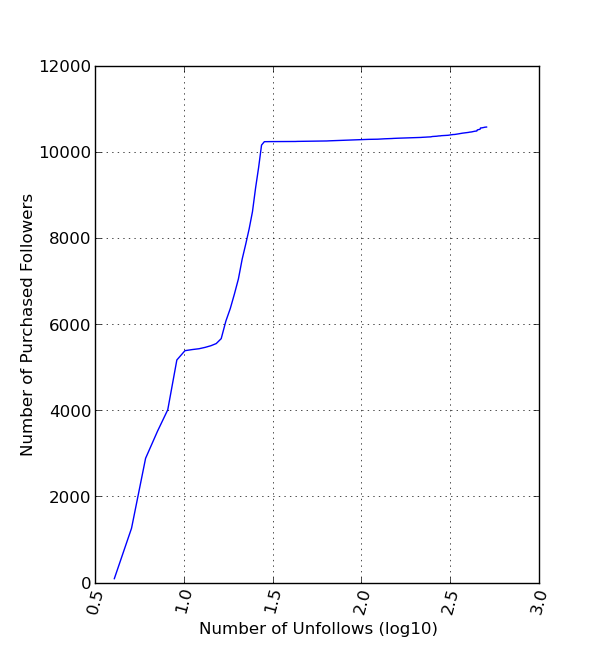}
    \caption{Unfollow Entropy Rate for Purchased Followers}\label{fig:unfollow-entropy}
\end{figure}

Figure~\ref{fig:unfollow-entropy} shows that a large fraction of purchased followers have a high unfollow entropy. The normalized entropy rate for 23\% purchased followers is as high as 0.76 and only 8\% users have a normalized unfollow entropy less than 0.21. To find out whether the users with higher unfollow entropy have lower quality than other users, we compared their normalized unfollow entropy rate with \emph{Klout} score. We found a strong negative correlation (\emph{Pearson correlation coefficient = -0.73}) indicating that users with higher unfollow entropy rate have low social reputation.  

\subsection{Social Engagement with Friends}
\label{subsec:socialengagement}
In this section, we explore how the purchased follower accounts are connected with their friends. We measure social engagement of users with their friends in form of retweets, @-mentions and favorite count. We also find out language overlap patterns between the users and his friends. 
\subsubsection{RTs and @-mentions}
We observed in section~\ref{subsub:inactive} that a large fraction of purchased accounts post only retweets instead of original content. We further explore whether these users retweet the content of their friends or not. If $RT_{count_i}$ is the number of tweets the user has retweeted of his friend $u_i$ and he has $N$ friends, then we define 
$$ Retweet Ratio = \frac{\frac{RT_{count_i}}{\Sigma_{i=1}^N RT_{count_i}}}{N * RT_{total}}$$ where $RT_{total}$ is the total number of retweets done by the user. This \emph{Retweet Ratio} quantifies the number of friends a user has retweeted and the number of times he retweeted them.

\begin{figure}[h!]
  \centering
    \includegraphics[width=0.4\textwidth]{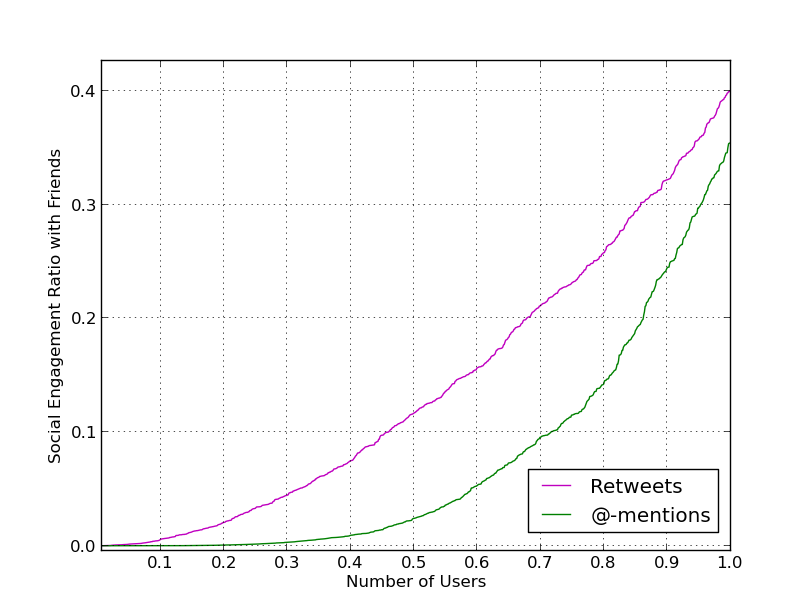}
    \caption{Social engagement of Purchased Users with their Friends}\label{fig:social_engagement}
\end{figure} 

Similarly we define the @-mention ratio to determine whether the user engages in conversations with his friends and to what extent.
$$ @mention Ratio = \frac{\frac{@_{count_i}}{\Sigma_{i=1}^N @_{count_i}}}{N * @_{total}}$$ where $@_{total}$ is the total number of @-mentions by the user. We observe in Figure~\ref{fig:social_engagement} that the highest Retweet Ratio score is 0.45 and the @-mention ratio is 0.35. This shows that though a large fraction of purchased accounts post only retweets, its not the tweets of their friends which they are retweeting. Similarly, low @-mention ratio suggests that purchased followers do not mention their friends. We found the maximum @-mention ratio with the followers of purchased users to be 0.32. This further strengthens our observation that purchased followers are low quality users and do not engage in conversations with their friends or followers.

\subsubsection{Language overlap with Friends and Followers}
\begin{figure}[h!]
  \centering
\subfigure[Distribution]{
        \label{subfig:language}
        \includegraphics[width=0.19\textwidth]{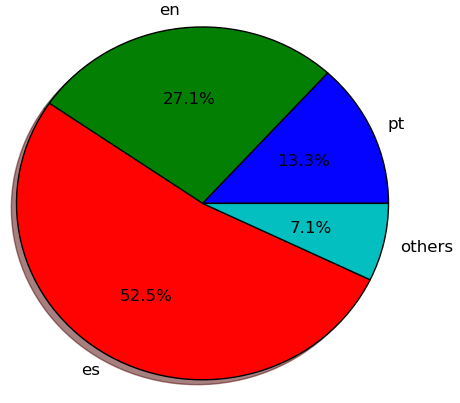} } 
        \subfigure[Number of Languages]{
        \label{subfig:NumLang}
        \includegraphics[width=0.24\textwidth]{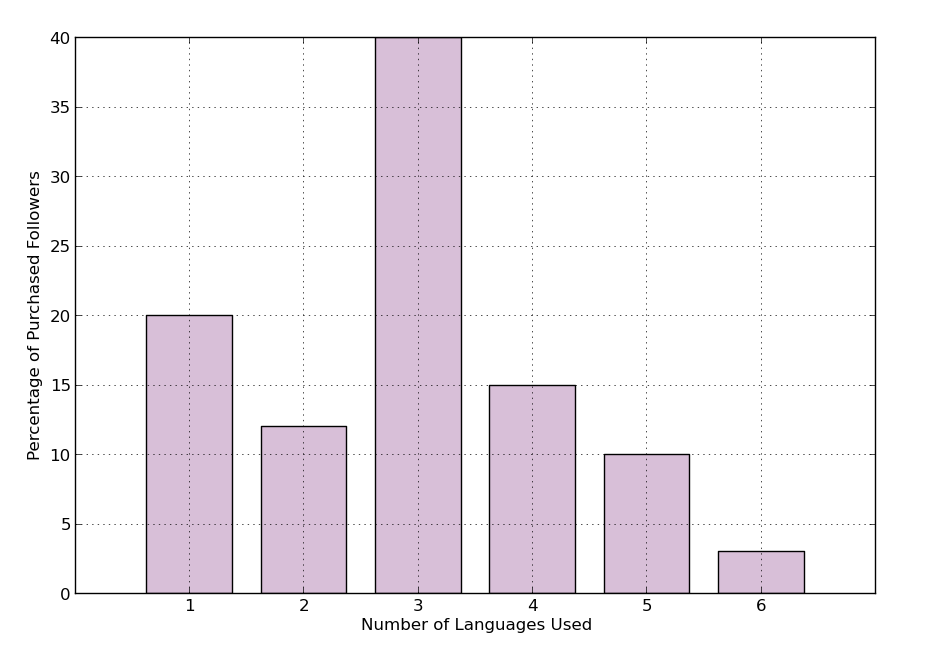} } 

    \caption{Languages used by Purchased Followers}\label{fig:languageAnalysis}
    
\end{figure}

We charecterize the language used by the purchased followers and the overlap with their friends. Figure~\ref{subfig:language} shows the distribution of language of purchased accounts. We observe that a 52\% of the users tweet in spanish. We also found that the purchased followers tweet and retweet in multiple languages as shown in Figure~\ref{subfig:NumLang}. Thirteen percent users used 5 or more languages. Only 32\% users posted tweets in less than or equal to two languages. We next find out the overlap of language amongst the purchased accounts with their followers and friends.

Users tweet and retweet in multiple languages. Therefore, we calculate the \emph{Language Overlap Score} for each user defined as

$$LangOverlap = \frac{\Sigma_{i=0}^N overlap_i}{N}$$ where N is the total number of friends or followers. If $L_f$ is the set of languages used by the friend/followers and $L_u$ is the set of languages used by the purchased user then $overlap_i$ with each friend/follower $u_i$ is defined as
\begin{equation*}
  overlap_i=\begin{cases}
    1, & \text{if $|L_f \cap L_u| \neq 0 $}.\\
    0, & \text{otherwise}.
  \end{cases}
\end{equation*}

\begin{figure}[h!]
  \centering
    \includegraphics[width=0.36\textwidth]{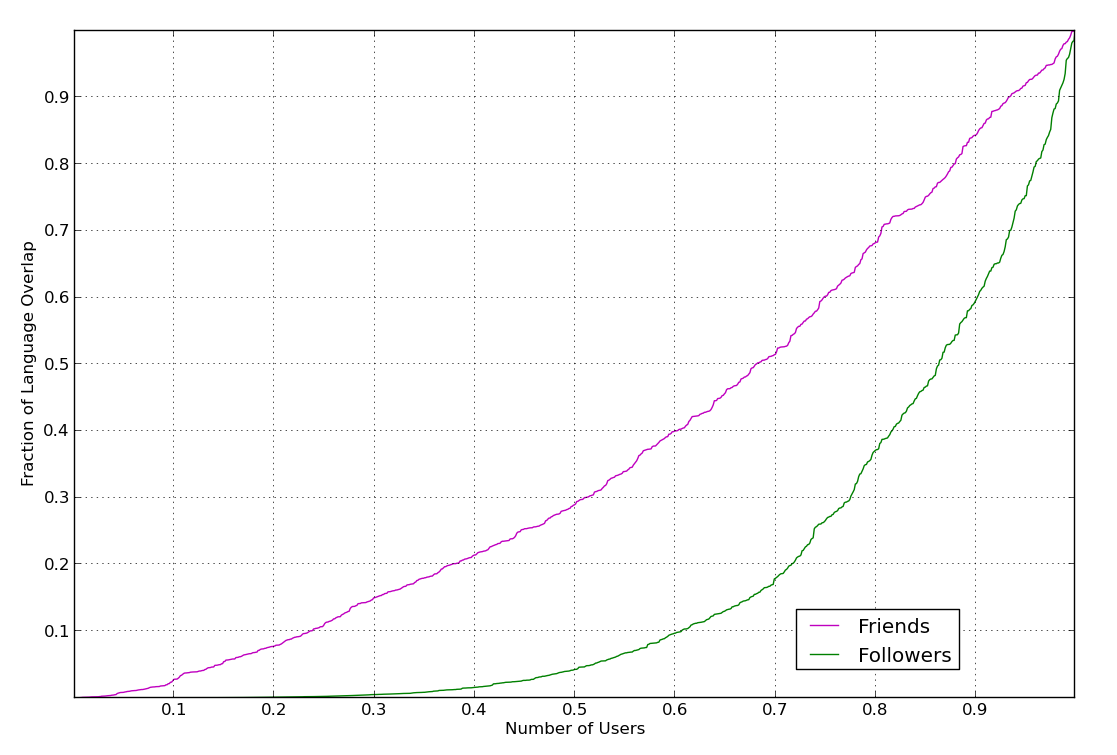}
    \caption{Language Overlap of Purchased Followers with their Friends and Followers}\label{fig:LanguageOverlap}
\end{figure}

We use the \emph{Language Overlap} score to determine how many users tweet in same language as their friends or followers. Figure~\ref{fig:LanguageOverlap} shows that 80\% users had an Overlap Score = 0.37 with their followers and Overlap Score = 0.68 with their friends. This indicates that a large fraction of purchased follower accounts do not care about the content posted by the users they are following. Also, the followers of these users do not have a high language overlap with them.  

\subsection{Spam Perpetrated by Purchased Follower Accounts}
In this section, we analyse the URLs posted by the purchased users. For each user, we collected his past 200 tweets and extracted the URLs if any. We observed that a large fraction of followers we purchased post tweets with URLs. Table~\ref{table:URLtweets} summarizes the number of tweets with URLs from each market we purchased followers.

\begin{table}[h!]
\caption{Tweets with URLs}\label{table:URLtweets}
\centering
\begin{tabular}{@{}llll@{}}
\toprule
Merchant & \begin{tabular}[c]{@{}l@{}}Public\\ User\end{tabular} & Tweets & \begin{tabular}[c]{@{}l@{}}Tweets\\ with URLs\end{tabular} \\ \midrule
buyfollowers & 902 & 83,936 & 45,945 \\
buy1000followers & 10,768 & 339,432 & 188,836 \\ \bottomrule
\end{tabular}
\end{table}

\begin{table*}[t!]
\centering
\caption{Spam URLs detected by blacklists}\label{table:spamURLs}
\begin{tabular}{@{}llllll@{}}
\toprule
Merchant & Spam URLs & PhishTank & \begin{tabular}[c]{@{}l@{}}Safebrowsing\\ API\end{tabular} & SURBL & VirusTotal \\ \midrule
buyfollowers & 2,504 & 200 & 1,856 & 1,710 & 1,021 \\
buy1000followers & 23,321 & 2,021 & 14,432 & 10,311 & 13,341 \\ \bottomrule
\end{tabular}
\end{table*}

To determine whether the URLs posted by the purchased followers are legitimate or not, we used multiple lookup services which maintain a blacklist of phishing, malware and other malicious URLs. We used PhishTank, Google Safebrowsing API, SURBL and VirusTotal API to lookup the URLs. We found that 12\% of the users we purchased posted one or more tweet with a URL blacklisted by one of the above services. We summarize the blacklist lookup results in Table~\ref{table:spamURLs}. We observed that 13.67\% tweets of purchased accounts were spam. Out of 234,781 tweets with URLs, we found that 32,117 tweets had a spam URL. The unique number of spam URLs were found to be 25,825.

\begin{figure}[h!]
  \centering
    \includegraphics[width=0.36\textwidth]{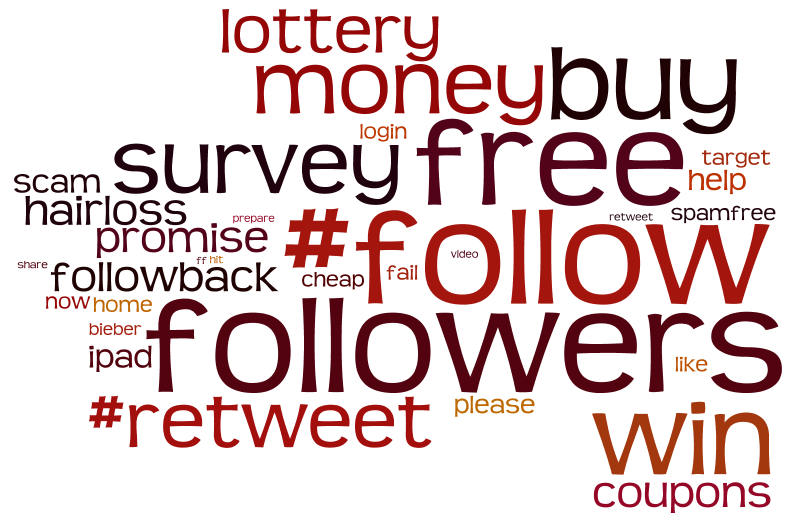}
    \caption{Word Cloud of Spam Tweets by Purchased Followers}\label{fig:spam_words}
\end{figure}

To understand the kind of spam users were perpetrating, we analyse content of the spam tweets which were in english. Figure~\ref{fig:spam_words} shows the most popular words which appear in tweets with spam URLs. We observed that a large fraction of spam was about the fake follower underground market. \emph{\#follow, followers, followback} keywords suggest that the purchased users were trying to spread propaganda about the follower market. The other kind of spam we observed was directed towards stealing credentials, i.e., phishing attack. Some of the keywords related to spam spread by the purchased followers were \emph{ipad, money, lottery}.

\section{Prediction of Suspicious Following Behaviour}
\label{sec:prediction}
In the second part of our study, we build a supervised predictive model to detect suspicious following behaviour on Twitter. In this section we explain the feature set used for the classification task and the experimental setup.

\subsection{Features for Classification}
For our prediction task to detect suspicious following behaviour we explore \emph{user profile}, \emph{network}, \emph{content} and \emph{user behaviour} based features. In all, we explore 18 features for our  classification task as described in Table~\ref{table:feats}. \emph{User profile} based features focus upon properties of the Twitter user profile information. The \emph{network} based features describe the relationship of the user with his friends and followers. We next explore the \emph{content} based features to understand the nature of tweets posted by the user and also investigate the \emph{behavioural} features to understand the tweeting patterns and follow dynamics exhibited by the user. For the \emph{network} based features, we constrain our analysis to single hop network of the users due to Twitter API rate limit restrictions. Also, we keep our \emph{content} based analysis limited to  stylistic features of tweets due to the presence of multi-lingual users in our dataset and the complexity of computation due to transliterated text, misspellings and use of short hand language. Table~\ref{table:feats} enlists all the feature sets we used for our prediction task. 

\begin{table}[h]
\small
\caption{Description of the feature sets}\label{table:feats}
\begin{tabular}{@{}llll@{}}
\toprule
Set & Category & Number & Features \\ \midrule
\multirow{4}{*}{A} & \multirow{4}{*}{User Profile} & \multirow{4}{*}{4} & presence of bio \\
\multicolumn{1}{c}{} &  &  & presence of URL in bio \\
\multicolumn{1}{c}{} &  &  & number of posts \\
\multicolumn{1}{c}{} &  &  & social reputation \\ \midrule
\multirow{2}{*}{B} & \multirow{2}{*}{Network} & \multirow{2}{*}{2} & follower / friends ratio \\
 &  &  & number of followers \\ \midrule
\multirow{6}{*}{C} & \multirow{6}{*}{Content} & \multirow{6}{*}{6} & hashtags per tweets \\
 &  &  & spam words used per tweet \\
 &  &  & length of tweet \\
 &  &  & number of languages used \\
 &  &  & number of RTs per tweet \\
 &  &  & @mentions per tweet \\ \midrule
\multirow{6}{*}{D} & \multirow{6}{*}{Behaviour} & \multirow{6}{*}{6} & unfollow entropy rate \\
 &  &  & RT engagement score \\
 &  &  & @mention engagement score \\
 &  &  & language overlap \\
 &  &  & time since last tweet \\
 &  &  & tweets per day \\ \midrule
\end{tabular}
\end{table}

We explained some of these features in the previous section; here we describe how we calculated the values of remaining features:
\vspace{-5mm}
\paragraph{Presence of bio and URL:} Some Twitter users give description about themselves on their profile which is called \emph{bio}. We check the presence of \emph{bio} for each user under inspection. We also check whether the user has mentioned any external URL in his \emph{bio} and use this as a feature.
\vspace{-5mm}
\paragraph{Social reputation:} We define social reputation by the \emph{Klout} score which gives an estimate of the impact score of the user on various online social media. 
\vspace{-5mm}
\paragraph{Hashtags per tweet:} We calculate the average number of hashtags used per tweet. We define this metric as
$$ \text{hashtag/tweet} = \frac{\Sigma_{tweet=0}^N \#hashtags}{\#tweets} $$
\vspace{-8.5mm}
\paragraph{Spam words used per tweet:} In the earlier section, we noticed that a fraction of purchased follower accounts also spread spam and malicious content. To detect spam in the tweet content, we use a spam word lookup list~\footnote{\url{http://www.mailup.com/spam-words-to-avoid.htm}} and define the following metric
$$ \text{spam words/tweet} = \frac{\Sigma_{tweet=0}^N \text{\#spam words}}{\#tweets} $$
\vspace{-8mm}
\paragraph{Time since last tweet:} We found that purchased followers exhibiting suspicious following behaviour have very less tweeting activity and a large fraction of such users are inactive. To measure  time since the account has been inactive, we find the difference in time in seconds since the latest tweet with the time of our experiment.

These are the discriminative features we use to distinguish between regular and suspicious following behaviour. With the help of these features, we detect users with suspicious follow behaviour in the following section. 

\subsection{Experimental Setup and Classification}
For our classification experiment, we consider the 11,760 public purchased followers as our true positive dataset of suspicious follow behaviour. For the negative class (legitimate follow behaviour), we pick random 11,760 users from Twitter stream using the streaming API. However, a balanced dataset as ours may create a sample bias. Therefore, to ensure valid results and eliminate the bias, we under-sample our negative class. We draw 10 random but independent subsets from the set of 11,760 legitimate users (-ve class) and train 10 classifier models based on these 10 subsets along with the 11,760 samples of the suspicious follow behaviour users (+ve class). We then use 10 fold cross validation and report the average results for our prediction task.

We treat the detection of suspicious follow behaviour as a two class classification problem. In order to detect such behaviour, we use several supervised learning algorithms like Naive Bayes, Gradient Decent, Random Forest etc. However, we achieved highest accuracy and overall best results with \emph{Support Vector Machine} (SVM). The goal of a SVM is to find the hyperplane that optimally separates the training data into two portions of an N-dimensional space where N is the total number of features used. A SVM performs classification by mapping input vectors into an N-dimensional space, and checking in which side of the defined hyperplane the point lies. We use a non-linear SVM with the Radial Basis Function (RBF) kernel for our experiment. Table~\ref{table:svm} gives the details of our experimental setup - dataset description and the parameter values for the SVM classification algorithm. 

\begin{table}[h]
\caption{Description of the experimental setup}\label{table:svm}
\small
\begin{tabular}{l|l}
\toprule
Dataset & 23,520 \\
`Suspicious' (+ve class) & 11,760 \\
`Legitimate' (-ve class) & 11,760 (10 times) \\
Classifier & SVM \\
C & 1,000 \\
alpha & 20.0 \\
Classification Runs & 10 \\
Feature Sets & \{A\}, \{A, B\}, \{A, B, C\}, \{A, B, C, D\} \\
Train-Test Split & 70\%-30\% \\
Cross Validation & 10-fold \\ \bottomrule
\end{tabular}
\end{table}

To reduce the error margin, we use a large C value for the RBF kernel of SVM. In order to assess the effectiveness of features, we repeat the classification experiment by incrementally adding each feature set. For evaluation, we used 70-30 split of the training and the testing dataset. We use 10 fold cross validation to report our results. 

\subsection{Classification Results and Evaluation}
Table~\ref{table:confusion} shows the confusion matrix for our classification task. The confusion matrix defines the percentage of false negatives and false positives. We were able to accurately classify 82.5\% users with suspicious follow behaviour and 88.3\% users with legitimate behaviour. This shows that we are able to detect suspicious following behaviour to a good extent. However, for the evaluation of our classification result, we used the standard evaluation metrics in this classification task -- accuracy, F-measure and Area under the Curve (AUC).
\begin{table}[h]
\center
\caption{Confusion Matrix -- Classification Results}\label{table:confusion}
\begin{tabular}{@{}cccc@{}}
\toprule
 &  & \multicolumn{2}{c}{Predicted} \\ \midrule
 &  & Suspicious & Legitimate \\
\multirow{2}{*}{True} & Suspicious & 82.5 & 17.5 \\
 & Legitimate & 11.7 & 88.3 \\  \bottomrule
\end{tabular}
\end{table}

\begin{figure}[h]
  \centering
    \includegraphics[width=0.45\textwidth]{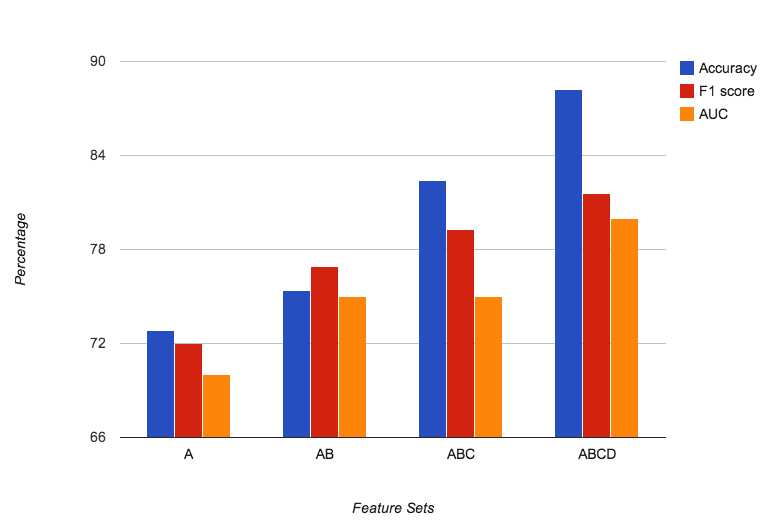}
\caption{Classification accuracy to predict suspicious following behaviour on incremental feature addition}\label{fig:incremental}
\end{figure}

As discussed in the previous section, we incrementally added feature sets to evaluate the effectiveness of all the features. Figure~\ref{fig:incremental} shows the performance of our classifier on Accuracy, F1 score and AUC metrics when feature sets are incrementally added. We see that each feature set has a positive effect on the performance of the classifier across all metrics. We also observed that adding behavioural based features suddenly increase the overall accuracy of our classification model. We received a maximum accuracy of 88.2\%.

\subsection{Feature Importance}
In this section, we look at the importance of features used for suspicious follow behaviour detection. We found that behavioural features play an important role in detecting suspicious behaviour. \emph{Unfollow entropy} rate plays an important role; it is defined as the frequency with which the user is unfollowing his friends over time. Some of the most informative features we received after our classification task were \emph{unfollow entropy, RT-engagement ratio, @mention-engagement ratio, Language Overlap and Social Reputation}. The other informative and discriminative features were the use of multiple hashtags and spam words in the tweets. The user profile based features were the least helpful in detection of suspicious follow behaviour. One possible reason for this could be that a large fraction of legitimate users do not add a \emph{bio} or engage in heavy conversations on Twitter.  

\section{Ethics}
We ensured that all money we paid to underground merchants to acquire fake followers was exclusively for Twitter accounts created and fully controlled by us and for the sole purpose of conducting experiments in this paper. We adhered to Twitter guidelines and did not contact any Twitter user or acquire his/her account credentials. We ensured that no Twitter user was harmed or benefitted as a result of this research experiment. This experiment was purely for research; we do not encourage users to purchase Twitter followers.

\section{Conclusion and Future Work}
\label{sec:future}
In this study we explored the dynamics of purchased follower accounts. We found some characteristic features of users which exhibit suspicious follow behaviour. We investigated the behavioural features of the followers purchased from underground Twitter follower market and found that a large fraction of users feep following and unfollowing their friends at regular basis - an activity which is unusual for a legitimate account holder. We thus define the term \emph{unfollow entropy} to measure the rate of \emph{unfollow} over time. In order to understand the dynamics of purchased follower accounts, we divided our study into two parts. In the first part, we studied the properties of users with suspicious follow activity and how they are different from regular Twitter users. In the next part, based on the discriminative features, we used supervised learning methodology to detect suspicious follow behaviour from regular behaviour. We received an overall accuracy of 88.2\%. 

In this study we only looked at one of the Twitter follower market schemes where there is no need to follow back the merchant or provide the Twitter password. The dynamics and network structure of such a market which requires password might be different from the one we focussed on in this study. In future we plan to compare the various markets and automatically detect merchants and customers to reduce this fraudulent activity on Twitter.

\section{Acknowledgement}
We thank all the members of Precog research group at IIIT-Delhi for their valuable feedback and support throughout this work. We also thank members of CERC at IIIT-Delhi for their encouragement and insightful comments.

\bibliographystyle{abbrv}
\small
\bibliography{sigproc}  
\nocite{*}
\balance
\end{document}